\documentstyle[prl,aps]{revtex}
\begin{document}


\title{Ice XII in its second regime of metastability}
\author{Michael Marek Koza}
\address{Fachbereich Physik, Universit\"at Dortmund, D-44221 Dortmund, Germany}
\author{Helmut Schober, Thomas Hansen}
\address{Institut Laue-Langevin, F-38042 Grenoble, France}
\author{Albert T\"olle}
\address{Fachbereich Physik, Universit\"at Dortmund, D-44221 Dortmund, Germany}
\author{Franz Fujara}
\address{TU-Darmstadt, Hochschulstr. 6, D-64289 Darmstadt}
\date{\today}
\maketitle

\begin{abstract}
We present neutron powder diffraction results which give unambiguous evidence
for the formation of the recently identified new crystalline ice phase
\cite{lobban}, labeled ice~XII, at completely different conditions.
Ice~XII is produced here by compressing hexagonal ice~I$_{\rm{h}}$
at T = 77, 100, 140 and 160\,K up to 1.8\,GPa.
It can be maintained at ambient pressure in the temperature
range 1.5 $<$ T $<$ 135\,K.
High resolution diffraction is carried out at T = 1.5\,K
and ambient pressure on ice~XII and accurate structural properties
are obtained from Rietveld refinement. 
At T = 140 and 160\,K additionally ice ~III/IX is formed.
The increasing amount of ice~III/IX with increasing temperature
gives an upper limit of T $\approx$ 150\,K for the successful formation
of ice~XII with the presented procedure.

\end{abstract}


Although, water has been extensively studied both experimentally and
theoretically it still rewards us with new and unexpected properties.
This has been demonstrated recently by (i) the detection of polyamorphism
\cite{mishima1,mishima2,mishima3} and (ii) the discovery of a new crystalline
phase (ice~XII) \cite{lobban,OKeeffe}.
Having been observed in different regions of water's phase diagram the two
phenomena were originally thought to be disconnected (Figure \ref{fig1}).
The phenomenon of polyamorphism, i.e. the existence of two distinct amorphous
phases, is still lacking a comprehensive understanding \cite{poole}.
Although polyamorphism is equally observed in other substances
particularly interesting explanations have been put forward for water.
These ideas are based on computer simulations \cite{mishima4} and link
polyamorphism to particularities of the supercooled liquid like phase
segregation and a second critical point.
Due to homogeneous crystallization supercooled water is
experimentally inaccessible in the region where these phenomena are
expected.
The hypotheses must, therefore, be checked indirectly, e.g. by
establishing the glassy character of the amorphous phases.
The formation of the high--density amorphous ice (HDA) ---
achieved by compressing crystalline hexagonal ice~I$_{\rm h}$
at temperatures below 150\,K to pressures exceeding 1\,GPa (Fig. \ref{fig1}) 
--- has received particular attention in this context
\cite{mishima4,mishima5,tse}.

So far, the formation of HDA from ice~I$_{\rm h}$ has been reported
as a well-defined transition channel.
The contamination of the amorphous samples by crystalline
impurities has been granted little attention \cite{bosio,bizid,bellissent}.
Only recently \cite{koza1,koza2,schober} strong experimental indications
have become available which imply that all these contaminations
correspond to ice~XII.
As ice~XII was originally observed in a completely
different region of water's phase diagram this shows
that it is a rather prolific phase of water.
Moreover, the co-production of ice~XII has, as we will argue in the
conclusions, far reaching implications for the I$_{\rm h}$ 
to HDA transformation
and, thus, on the origin of water's polyamorphism.

In this letter we show that the structure of ice~XII produced at low 
temperatures is definitely identical with the phase characterized 
by Lobban et al. \cite{lobban} at higher temperatures.
Furthermore, we find that no continuous connection between the two 
regions of apparent metastability exists.
And finally, we identify the conditions which define whether the 
compression of I$_{\rm{h}}$ results in ice~XII or HDA.

Our results are based on high--resolution neutron powder
diffraction experiments on samples which are produced at various temperatures.
All samples are treated in the following way.
About 2.5 ml of D$_2$O (purity 99.9, resistivity 1 M$\Omega$cm)
is frozen to common hexagonal ice~I$_{\rm{h}}$ and cooled to 77\,K in a
piston--cylinder apparatus.
The samples are heated to the desired temperatures, namely 77, 100, 140 and
160\,K
and tempered for about 30 minutes.
Each sample is then compressed to a maximum nominal pressure of 1.8\,GPa.
The rate of compression is 1\,GPa/min at 77\,K and 0.5\,GPa/min at
all other temperatures.
Once the maximum pressure is attained the samples are cooled back to 77\,K
and finally recovered from the pressure device in liquid nitrogen,
where they are powdered by using a mortar and a pestle.
No pressure changes which could be interpreted as signs of phase transitions
are observed in the course of cooling under pressure.
The sample temperature as measured at the bottom of the pressure cylinder
apparatus does not increase by more than 10\,K during the compression
phase.

The diffraction experiments are carried out on the high--resolution
diffractometer D2B at the Institute--Laue--Langevin in Grenoble,
France \cite{yellowbook}.
A Vanadium sample holder (diameter = 7 mm) and a standard cryostat
are chosen as sample environment \cite{yellowbook}.
Two different experimental setups are applied.
The accurate structure determination of ice~XII is carried out
at 1.5\,K on the sample produced at 77\,K.
For this measurement the horizontal incident beam divergence $\alpha_1$ is set
to 10' and the monochromator aperture (MA) to 10 mm.
The detector is moved using $\Delta\Theta = 0.05^{\circ}$ steps.
In this high resolution mode, the applied wavelength
$\lambda = 1.59427 \pm 0.00006$ \AA\
and the instrumental resolution are determined from an independent measurement
of a silicon powder sample.
For good statistics data are collected over 15 hours.
All other measurements are performed at 110\,K, with $\alpha_1$ = 30'
and MA = 50 mm, $\Delta\Theta = 0.05^{\circ}$ at the given wavelength.
These lower resolution data are collected for 120 minutes
(sample prepared at 77\,K)
and 30 minutes (all other samples) which is sufficient for a structural
identification of the samples.

Figure \ref{fig2} displays the results of the high--resolution measurement
at 1.5\,K.
A Rietveld refinement \cite{rietveld} is performed without any constrains 
on the parameters using the program Fullprof \cite{fullprof}.
Our refinement confirms that the structure of the observed ice phase
is unambiguously that of ice~XII \cite{lobban,koza1}.
This structure is characterized by twelve water molecules arranged
in a tetragonal unit cell meeting the symmetry space group
I\={4}2d.
The refined unit cell parameters are $a = 8.2816 \pm 0.0002$\,\AA\ and
$b = 4.0361 \pm 0.0001$\,\AA\ which result in a calculated microscopic
density of $\rho = 1.4397 \pm 0.0003$\,g/cm$^3$ (D$_2$O).
We present in Table \ref{tab1} the refined fractional coordinates and thermal
factors and in Table \ref{tab2} the calculated intra-- and intermolecular
distances and bond angles.
Taking into account the different preparational conditions which are
used here and in reference \cite{lobban} the cell constants, mass densities
and atomic parameters are in good agreement with each other.
Profile features in the diffraction pattern (Figure~\ref{fig2})
which are not due to ice~XII can be attributed to the sample environment
and a slight contamination of the sample by some untransformed I$_{\rm{h}}$
and simultaneously produced amorphous ice.
This contamination is taken into account in the refinement 
by profile (I$_{\rm{h}}$, symmetry space group P$6_3$/mmc) 
and background (amorphous ice) matching
\cite{fullprof,pawley}.
The peak which due to the sample environment is excluded from
the calculation.
The formation of amorphous ice shows the ambivalent nature 
of the applied pressure induced transition.
The final residuals of the refinement are $R_{wp}=3.48 \%$ and
$R_p=2.69 \%$.

The diffraction results for samples prepared at different temperatures are
compared in Figure \ref{fig3}.
The presence of ice~XII can be clearly identified in all of the data,
although,
its diffraction pattern is obscured progressively by an additional
contribution at 140 and 160\,K, respectively.
To identify this additional contribution profile matching methods
\cite{fullprof,pawley}
are used for ice phases expected in this region of water's phase
diagram.
The most promising indexing scheme is given by the symmetry space group
P$4_12_12$.
This symmetry space group is uniquely inherent to ice~III and ice~IX, 
which differ in the degree of order in their proton sublattices
\cite{kamb,arnold,laplaca,londono}.
As in ice~XII, twelve water molecules inhabit the tetragonal unit cell of
ice~III/IX.
Table \ref{tab3} gives the unit cell constants and the corresponding
mass densities of ice~XII and ice~III/IX as determined by the profile
matching.
The profile shape function, which is used in the matching procedure,
is fully accounted for by the instrument resolution function as determined
from the samples prepared at 77 and 100\,K (Figure 3a,b).
As a consequence only the cell constants (Table \ref{tab3}) are freely
adjustable for the phase mixtures.
Typical residuals of the matching for the ice~XII and ice ~III/IX mixtures
are $R_{wp}\approx 6 \%$ and $R_p\approx 8 \%$.
The inset of Figure \ref{fig3} shows a diffraction pattern taken at
$\lambda=3.00$\,\AA\ on the time--of--flight spectrometer IN5
of a sample prepared at $T\approx 165$\,K.
The absence of the (220) peak of ice~XII which should be well resolved at
$2\Theta \approx 70^{\circ}$ indicates that the content of ice~XII 
is negligibly small.
Thus, above $T > 160$\,K solely ice~III/IX is formed.
The pressure induced formation of ice~III/IX from I$_{\rm{h}}$
at temperatures exceeding 140\,K is in agreement with recent extensive 
studies \cite{garg,mishima4,mishima5}.
Formation of ice V at temperatures exceeding 180\,K \cite{mishima5}
is not observed in any of our samples. 
This shows that despite the elevated compression rates used in our work the
temperature of the sample during the compression stage stayed close to the
cell temperature.

Summarizing, we have demonstrated that ice~I$_{\rm{h}}$ can
be successfully compacted either into the high--density amorphous state (HDA)
or into the crystalline phase ice~XII.
Both phases are produced  by application of pressure exceeding 1.0\,GPa at
temperatures below 150\,K (Figure \ref{fig1}).
Ice~XII is a rather prolific feature in waters phase
diagram.
Its formation in two seemingly disconnected regions is
unusual but not entirely surprising when we consider the
fact that the formation is governed by dynamic variables
like the compression rate, and, in addition, is in competition
with other crystalline and amorphous phases.
Ice~XII can be recovered at ambient pressure and low temperature and
can be stored at temperatures lower than 135\,K.
At higher temperatures it starts to transform apparently to the
metastable cubic phase \cite{koza1,kuhs} which itself is a precursor of the
stable I$_{\rm{h}}$ form.
Production of different ice phases under seemingly identical
conditions is not a new observation but related to the non-equilibrium
character of the transitions.
In the case of the metastable ice phases IV, XII and stable ice V
(see Figure \ref{fig1}) phase discrimination is achieved via the cooling
rates \cite{lobban,kamb2,engelhardt1,engelhardt2}.
In our example the compression rate seems to be the decisive control
parameter.
Having used higher compression rates than in previous experiments
it was possible to obtain exclusive formation of ice~XII.
Given the higher density of recovered ice~XII ($\rho\approx 1.44$\,g/cm$^3$)
in comparison to recovered HDA ($\rho\approx 1.30$\,g/cm$^3$) and an
anticipated kinetic character of the transitions such a dependence could be
expected.
The observation of explosive sound accompanied by abrupt loss of pressure
indicates
the development of shock waves during the compression which could play a major
role in the transformation process.

The here established competition between crystallization and amorphization
under close experimental conditions has to be properly acknowledged by all
theoretical attempts trying to explain amorphous ice formation under pressure.
Crystallization implies a reorganization and not merely a deformation
of water's hydrogen bond network and, therefore, places more stringent
conditions on the transition mechanism.
A purely mechanical instability \cite{tse} as recently proposed for the
formation of HDA
is, to our opinion, insufficient to explain the ice~XII formation.
This holds unless the mechanical collapse is accompanied by high molecular
mobility
as for example in the case of a thermodynamic mechanical instability
\cite{shpakov}
or shock wave melting \cite{kormer} which allow for crystalline reassembly.
The requirement for high mobility, necessary for ice XII formation, equally
explains the threshold value of around 1.0 GPa which corresponds to  water's
extrapolated melting line.

\acknowledgements
Helpful discussions with W.F. Kuhs  are greatfully acknowledged.
This work is financially supported by the German {\it Bundesministerium f\"ur
Bildung und Forschung} project No. 03--FU4DOR--5.



\begin{table}
\caption{\bf Fractional coordinates x, y, z, multiplicities M of atomic
positions
and corresponding thermal factors B$_{iso}$ of ice~XII (prepared at 77\,K
and measured
at 1.5\,K and ambient pressure) as obtained from Rietveld refinement.
Errors are given in standard Fullprof units.}
\vskip3mm
\begin{tabular}{cccccc}
 & x & y & z & M & B$_{iso}$ \\\hline
O(1) & 0 & 0 & 0 & 4 & 0.437(39) \\
O(2) & 0.36464(20) & 0.25 & 0.125 & 8 & 0.905(28) \\
D(3) & 0.04044(32) & 0.08375(24) & -0.14212(78) & 16 & 1.512(45) \\
D(4) & 0.28954(28) & 0.21988(31) & 0.29498(74) & 16 & 1.840(55) \\
D(5) & 0.42120(29) & 0.33618(26) & 0.23683(86) & 16 & 1.522(34) \\
\label{tab1}
\end{tabular}
\end{table}
\vskip0cm
\begin{table}
\caption{\bf Intra-- and intermolecular distances, bond lengths and their
multiplicities
for water molecules  of ice~XII determined at 1.5\,K and ambient pressure. }
\vskip3mm
\begin{tabular}{ccc}
Bond & Distance (\AA) & Multiplicity \\ \hline
O(1)--O(2) & 2.80 & 6 \\
O(2)--O(2) & 2.77 & 2 \\
O(1)--D(3) & 0.96 & 4 \\
O(2)--D(4) & 0.96 & 2 \\
O(2)--D(5) & 0.97 & 2 \\ \hline
& & \\
Bond angle & Angle ($\circ$) & Multiplicity \\ \hline
O(2)--O(1)--O(2) & 107.0 & 4\\
O(2)--O(1)--O(2) & 114.5 & 2\\
O(2)--O(2)--O(2) & 93.5 & 1\\
O(2)--O(2)--O(1) & 131.9 & 2\\
O(2)--O(2)--O(1) & 83.1 & 2\\
O(1)--O(2)--O(1) & 132.8 & 1\\
D(3)--O(1)--D(3) & 110.9 & 4 \\
D(3)--O(1)--D(3) & 106.6 & 2 \\
D(4)--O(2)--D(4) & 99.1 & 1 \\
D(4)--O(2)--D(5) & 117.2 & 2 \\
D(4)--O(2)--D(5) & 99.9 & 2 \\
D(5)--O(2)--D(5) & 122.0 & 1 \\
\label{tab2}
\end{tabular}
\end{table}
\vskip1cm
\begin{table}
\caption{\bf Cell constants a, c and mass density $\rho$
(D$_2$O) of ice~XII and ice ~III/IX in samples prepared
at given temperatures T and measured at T = 110\,K as
determined by a profile matching method.
}
\vskip3mm
\begin{tabular}{c|ccc|ccc}
& \multicolumn{3}{c}{ice~XII} & \multicolumn{3}{c}{ice ~III/IX} \\ \hline
 T (K) & a (\AA) & c (\AA) & $\rho$ (g/cm$^3$) & a (\AA) & c (\AA) & $\rho$
(g/cm$^3$) \\ \hline
 77  & 8.30 & 4.04 & 1.43 & -- & -- & -- \\
 100 & 8.30 & 4.04 & 1.43 & -- & -- & -- \\
 140 & 8.30 & 4.04 & 1.43 & 6.71 & 6.99 & 1.27 \\
 160 & 8.30 & 4.03 & 1.44 & 6.71 & 6.99 & 1.27
\label{tab3}
\end{tabular}
\end{table}

%
%
%
%
%
%
%
%
\begin{figure}
\caption{The phase diagram of water.
($\bullet$) represents the region in which ice~XII is observed
by Lobban et al. ($T = 260$\,K, $p = 0.55$\,GPa).
Please note that this region is fully encapsulated in the regime
of the stable ice V.
In addition, a third water phase namely the metastable ice~IV can
equally be formed in this region by employing an appropriate cooling process
which is indicated by the vertical arrow.
The insert sketches the pressure induced transition line of I$_{\rm{h}}$
(thick solid line) as studied by O. Mishima and followed up here.
The light shaded area stresses the region in which ice~XII
is successfully formed in the present work.
Horizontal arrows indicate that I$_{\rm{h}}$ transforms by compression
below 150\,K to ice~XII or HDA and above 150\,K to ice ~III/IX.
The dark shaded area displays the 150\,K boundary studied by O. Mishima
for HDA and equally observed for ice~XII here.}
\label{fig1}
\end{figure}
\begin{figure}
\caption{Neutron powder diffraction patterns of ice~XII
measured (bars) on the high--resolution diffractometer D2B
($\lambda=1.594 $ \AA) at 1.5\,K and ambient pressure
and calculated (solid line) from a Rietveld refinement using
the package Fullprof.
Tick marks indicate the reflection positions of ice~XII (upper panel)
and untransformed I$_{\rm{h}}$ (lower panel, relative amount of 0.8 \%).
The lower solid line gives the difference profile between
the measured and calculated diffraction pattern.
The arrow indicates a profile feature arising from the sample environment
which
is excluded from the refinement.
The inset gives an enlarged plot of the data at $58^{\circ} \ge 2\Theta \ge
150 ^{\circ}$.}
\label{fig2}
\end{figure}
\begin{figure}
\caption{Diffraction patterns of samples prepared at different temperatures
and measured with $\lambda=1.594$ \AA\ at T = 110\,K and ambient pressure.
The samples are produced at 77\,K (a), 100\,K (b), 140\,K (c) and 160\,K (d).
>From profile matching calculations the patterns can be accounted for by
ice~XII (a, b) only and mixtures of ice~XII and ice ~III/IX (c, d).
Note that the amount of ice ~III/IX is progressively increasing as the
temperature is raised.
The inset shows a diffraction pattern taken at the time--of--flight
spectrometer
IN5 at $\lambda=3.00$ \AA\  of a sample prepared at $T\approx 165$\,K.}
\label{fig3}
\end{figure}

\begin{references}

\bibitem{mishima1} O. Mishima, L.D. Calvert and E. Whalley, {\it Nature},
{\bf 310}, 393, (1984).

\bibitem{mishima2}  O. Mishima, L.D. Calvert and E. Whalley, {\it Nature},
{\bf 314}, 76, (1985).

\bibitem{mishima3} O. Mishima, {\it J. Chem. Phys.}, {\bf 98}, 4878, (1993).

\bibitem{lobban} C. Lobban, J. L. Finney and W. F. Kuhs,
                 {\it Nature}, {\bf 391}, 268, (1998).

\bibitem{OKeeffe} M. O'Keeffe, {\it Nature}, {\bf 392}, 879, (1998).

\bibitem{poole} P.H. Poole, T. Grande, C.A. Angell and P.F. McMillan,
		{\it Science}, {\bf 275}, 322, (1997).

\bibitem{mishima4} O.Mishima and H.E. Stanley,
		   {\it Nature}, {\bf 396}, 329, (1998).

\bibitem{mishima5} O. Mishima, {\it Nature}, {\bf 384}, 546, (1996).

\bibitem{tse} J.S. Tse, D.D. Klug, C.A. Tulk, I. Swainson, E.C.
Svensson, C.--K. Loong, V. Shpakov, V.R. Belosludov, R.V. Belosludov
and Y. Kawazoe, {\it Nature}, {\bf 400}, 647, (1999).

\bibitem{bosio} L. Bosio, G.P. Johari and J. Teixeira , {\it Phys. Rev.
Lett.},
{\bf 56}, 460, (1986).

\bibitem{bizid}  A. Bizid, L. Bosio, A. Defrain and M. Oumezzine,
{\it J. Chem. Phys.}, {\bf 87}, 2225, (1987).

\bibitem{bellissent} M.-C. Bellissent-Funel, J. Teixeira  and L. Bosio,
{\it J. Chem. Phys.}, {\bf 87}, 2231, (1987).

\bibitem{koza1} M. Koza, H. Schober, A. T\"olle, F. Fujara, and T. Hansen,
{\it Nature}, {\bf 397}, 660, (1999).

\bibitem{koza2} M.M. Koza, ILL Report No.ILL97KO10T, {\it Report on the
preparation and
performance of time--of--flight experiments on the amorphous and
polycrystalline solid D$_2$O}, p. 32 -- 36, (1997).

\bibitem{schober} H. Schober, M. Koza, A. T\"olle, F. Fujara, C.A.
Angell and R. B\"ohmer, {\it Physica B}, {\bf 241--243}, 897, (1998).

\bibitem{yellowbook} The Yellow Book -- {\it Guide to neutron research
facilities at the ILL}, H.G. B\"uttner, E. Lelievre--Berna and F. Pinet,
Institut Laue--Langevin, 1997.

\bibitem{rietveld} H.M. Rietveld, {\it J. Appl. Cryst.}, {\bf 2}, 65, (1969).

\bibitem{fullprof} J. Rodriguez-Carvajal, {\it Rietveld, Profile Matching
and Integrated Intensities Refinement of X-ray and/or
Neutron Data},  Laboratoire Leon Brillouin (CEA-CNRS).

\bibitem{pawley} G.S. Pawley, {\it J. Appl. Cryst.}, {\bf 14}, 357, (1981).

\bibitem{kamb} B. Kamb and A. Prakash, {\it Acta Cryst.}, {\bf B 24}, 1317,
(1968).

\bibitem{arnold} G.P. Arnold, R.G. Wenzel, S.W. Rabideau, N.G. Nereson and
A.L. Bowman,
{\it J. Chem. Phys.}, {\bf 55}, 589, (1971).

\bibitem{laplaca} S.J. La Placa and W.C. Hamilton, {\it J. Chem. Phys.},
{\bf 58}, 567, (1973).

\bibitem{londono} J.D. Londono, W.F. Kuhs and J.L. Finney,
{\it J. Chem. Phys.}, {\bf 98}, 4878, (1993).

\bibitem{garg} A.K. Garg, {\it phys. stat. sol. (a)}, {\bf 110}, 467, (1988).

\bibitem{kuhs} W.F. Kuhs, D.V. Bliss and J.L. Finney, {\it J. Physique},
{\bf 48},
C--1631, (1987).

\bibitem{kamb2}  B. Kamb, A. Prakash and K. Knobler, {\it Acta Cryst.},
{\bf 22}, 706, (1967).

\bibitem{engelhardt1} H. Engelhardt and E. Whalley, {\it J. Chem. Phys.},
{\bf 56}, 2678, (1972).

\bibitem{engelhardt2} H. Engelhardt and B. Kamb, {\it J. Chem. Phys.}, {\bf
75}, 5887, (1981).

\bibitem{shpakov} V.P. Shpakov, J.S. Tse, V.R. Belosludov and R.V.
Belosludov, {\it J. Phys. Condens. Matter}, {\bf 9}, 5853, (1997).

\bibitem{kormer} S.B. Kormer, {\it Sov. Phys. Uspekhi}, {\bf 11}, 229, (1968).

\end{references}
\end{document}